\documentclass[letterpaper,twoside,journal]{IEEEtran}
\usepackage{amsmath,amsfonts}
\usepackage{algorithmic}
\usepackage{algorithm}
\usepackage{array}
\usepackage[caption=false,font=normalsize,labelfont=sf,textfont=sf]{subfig}
\usepackage{textcomp}
\usepackage{stfloats}
\usepackage{url}
\usepackage{verbatim}
\usepackage{graphicx}
\usepackage{cite}
\usepackage[capitalise]{cleveref}
\usepackage{aliascnt}
\crefname{section}{Section}{Sections} 
\newaliascnt{appendixsection}{section}
\crefname{appendixsection}{Appendix}{Appendices}
\aliascntresetthe{appendixsection}

\usepackage{tcolorbox} 
\newtcolorbox[auto counter]{mybox}[2][]{%
  colback=gray!10!white, colframe=black,
  fonttitle=\bfseries,
  title=Box~\thetcbcounter: #2,
  label=#1
}
\crefname{tcb@cnt@mybox}{Box}{Boxes}

\begin{document}

\title{Analytical bounds for decoy-state quantum key distribution with discrete phase randomization}
\author{Zhaohui Liu,
  \and
  Ahmed Lawey,
  \and
  and Mohsen Razavi
  \thanks{This work was supported in part by the UK EPSRC Grants EP/Y037421/1 and EP/X040518/1, and the Doctoral Scholarship offered by School of Electronic and Electrical Engineering at the University of Leeds.}
  \thanks{Authors are with the School of Electronic and Electrical Engineering, University of Leeds, Leeds LS2 9JT, United Kingdom (e-mail:ml20z8l@leeds.ac.uk; A.Q.Lawey@leeds.ac.uk; M.Razavi@leeds.ac.uk).}
}



\maketitle

\markboth{IEEE Journal on Selected Areas in Communications}
{Liu \MakeLowercase{\textit{et al.}}: Analytical Bounds for Decoy-State QKD with Discrete Phase Randomization}

\begin{abstract}
We analyze the performance of quantum key distribution (QKD) protocols that rely on discrete phase randomization (DPR). For many QKD protocols that rely on weak coherent pulses (WCPs), continuous phase randomization is assumed, which simplifies the security proofs for such protocols. However, it is challenging to achieve such a perfect phase randomization in practice. As an alternative, we can select a discrete set of global phase values for WCPs, but we need to redo the security analysis for such a source. While security proofs incorporating DPR have been established for several QKD protocols, they often rely on computationally intensive numerical optimizations. To address this issue, in this study, we derive analytical bounds on the secret key generation rate of BB84 and measurement-device-independent QKD protocols in the DPR setting. Our analytical bounds closely match the results obtained from more cumbersome numerical methods in the regions of interest.
\end{abstract}


\maketitle

\section{Introduction}
\par Quantum key distribution (QKD) aims to establish a secure communication channel between two distant parties, conventionally referred to as Alice and Bob. The theoretical security of QKD protocols is based on quantum mechanics \cite{lo1999unconditional,shor2000simple}, as well as a set of assumptions on the behavior and characteristics of the physical devices involved \cite{hoi-kwong2005decoy}. In real-world scenarios, these assumptions are not necessarily fully met due to a mismatch between the models assumed in the security analysis and that of the actual hardware, leading to potential vulnerabilities \cite{abellan2014ultra,tang2014experimental,pathak2023phase}. Consequently, accurate and comprehensive modeling of devices used is essential for conducting realistic security analyses and ensuring the robustness of QKD implementations. In this work, we focus on the specific issue of phase randomization when weak coherent pulses (WCPs) are employed in QKD protocols. In particular, we look at implementations that rely on discrete phase randomization (DPR), as opposed to continuous phase randomization (CPR). While the latter is an easier case to analyze, it would be more practical to establish alignment between the theory and practice in the DPR case.

\par In many QKD systems, WCPs are widely adopted as a substitute for ideal single-photon sources \cite{nahar2023imperfect,yang2024high}, which require more advanced technology. WCPs, typically generated by attenuated laser sources, offer a practical and scalable solution due to their ease of generation, cost-effectiveness, and compatibility with existing optical communications infrastructure. A critical aspect of using WCPs securely lies in the application of phase randomization. By using CPR, the output state of the WCP-based encoder can be modeled as a statistical mixture of photon-number (Fock) states, with a Poisson distribution. This transformation is fundamental to the decoy-state method \cite{ma2005practical,lo2006security}, which often enables tight estimation of the single-photon components and protects against certain eavesdropping strategies, such as photon-number-splitting attacks \cite{huttner1995quantum,wang2005beating}. In the absence of phase randomization, the coherent states retain their phase coherence, invalidating the assumptions underpinning standard security proofs for the decoy state QKD and rendering the system susceptible to sophisticated quantum attacks \cite{tang2013source,curras2023security}. 

\par Despite theoretical advantages of CPR, achieving ideal CPR in practical QKD systems remains a challenging task. Passive phase randomization techniques that rely on turning the laser sources on and off to generate random global phase values have limitations on the pulse rate because of the transient time needed to stabilize the laser source. Relying on free-running lasers in conjunction with amplitude modulators could also result in correlated phase values for consecutive pulses \cite{xu2012ultrafast}, especially at high repetition rates, again not meeting the required conditions for CPR \cite{abellan2014ultra}. {For example, in twin-field QKD settings, active phase randomization is employed to ensure security against coherent attacks and to reach long distances \cite{wang2019beating,wang2022twin}.  In practice, however, these schemes are constrained by the precision and stability of the modulation hardware, making it challenging to realize full CPR \cite{Curras_PRApp2021}.} Such discrepancies between theoretical models and practical implementations can introduce subtle vulnerabilities in the security of QKD protocols.

\par As a pragmatic alternative, DPR has been proposed \cite{cao2015discrete}. In DPR-based implementations, the global phase of each pulse is selected from a finite, predefined set of discrete values. Although, in this approach, we have more control on the phase values used, hence can better ensure the match between theory and implementation, the source model becomes more complicated. Specifically, the resulting quantum channel no longer conforms to the standard photon-number channel model that underpins conventional decoy-state analysis \cite{hoi-kwong2005decoy,ma2005practical,sun2013practical,li2022improving}. This deviation invalidates the assumptions used in traditional security proofs \cite{hoi-kwong2005decoy}, necessitating the development of new theoretical frameworks and security analyses tailored to DPR cases. 

\par The challenge of incorporating DPR into QKD protocols has been addressed for both BB84 \cite{cao2015discrete} and measurement-device-independent (MDI) QKD \cite{cao2020discrete} protocols. However, a significant limitation of existing approaches is related to their parameter estimation, which typically requires solving complex numerical optimization problems. The computational burden of these methods increases substantially with the number of discrete phase slices, making them less practical for real-time implementations, or when there are restrictions on computational power and memory, which is the case in many IoT scenarios \cite{wang2023finite,nahar2023imperfect}. 

\par In this work, we present an analytical framework for deriving bounds on the secret key generation rate in both BB84 and MDI QKD protocols in DPR settings. Our approach circumvents the need for computationally intensive numerical optimization by providing closed-form expressions that are easy to calculate. We demonstrate that the analytical bounds closely approximate those obtained through numerical methods, particularly as the number of discrete phase slices increases. Moreover, the proposed analytical techniques are not limited to BB84 and MDI QKD protocols but are also applicable to a broader class of MDI-type protocols that share similar source characteristics. This generality enhances the practicality and adaptability of our methods for real-world QKD deployments.

\par The structure of this paper is as follows. In \cref{Protocol description}, we introduce the theoretical framework, including the distinctions between CPR and DPR, and their implications for BB84 and MDI-QKD protocols.  Building on the concepts from \cite{cao2015discrete}, we define the optimization problems for parameter estimation and explain the relevant constraints in \cref{Key rate analysis}. Our analytical solutions are then presented in \cref{Analytical estimation}. In \cref{Simulation}, the numerical and analytical techniques are compared in certain simulation scenarios. Finally, we conclude our paper in \cref{Summary}. 

\section{System Description \label{Protocol description}}
\par In this section, we first explain how the source is modeled in the CPR and DPR cases. We then describe BB84 and MDI QKD protocols with DPR in the particular case of vacuum+weak decoy states with {time-bin encoding in \(Z\) and \(X\) bases} \cite{ma2005practical,cao2015discrete,cao2020discrete}. 
\subsection{Source description \label{Source description}}
\par Many QKD protocols employ WCP sources due to their practical feasibility and compatibility with existing optical communications infrastructure. A WCP is characterized by a global phase \(\theta\) and an average photon number (or intensity) \(\mu\), and can be mathematically modeled as the following coherent state:
\begin{equation}
    \begin{split}
        |\sqrt{\mu}e^{i\theta}\rangle=e^{-\frac{\mu}{2}}\sum_{n=0}^\infty\frac{\sqrt{\mu}^ne^{in\theta}}{\sqrt{n!}}|n\rangle,
    \end{split}
    \label{eq:ProtocolDescription_coherent state with phase}
\end{equation}
where \(|n\rangle\), \(n=0,1,...\), represents the Fock state of $n$ photons in the optical mode defined by the pulse.

In the CPR case, we assume that \(\theta\) is uniformly distributed in \([0,2\pi)\). The state generated by the WCP source can then be expressed by the following density matrix
\begin{equation}
    \begin{split}
        \frac{1}{2\pi}\int^{2\pi}_0 |\sqrt{\mu}e^{i\theta}\rangle\langle\sqrt{\mu}e^{i\theta}|d\theta=e^{-\mu}\sum_{n=0}^\infty \frac{\mu^n}{n!}|n\rangle\langle n|,
    \end{split}
\end{equation}
which is diagonal in the Fock basis.
This model, known as the photon-number channel, is at the core of many security proofs, particularly those involving WCPs with decoy states \cite{ma2005practical,tupkary2025qkd}. However, realizing perfect phase randomization in practice is technologically demanding. To address this, DPR has been proposed as an alternative.  
\par In DPR, the global phase is uniformly selected from a discrete set of \(D\) values, \(\theta_j=2\pi j/D\), where \(j=0,\ldots,D-1\), \cite{cao2015discrete}. The source output can then be modeled by the following density matrix
\begin{align}
     \frac{1}{D}\sum_{j=0}^{D-1}|e^{i2\pi j/D}\sqrt{\mu}\rangle\langle e^{i2\pi j/D}\sqrt{\mu}| 
    = \sum_{k=0}^{D-1}p_k^\mu|\lambda_k\rangle\langle\lambda_k|, \label{eq:Source description_DPR channel}
\end{align}
where
\begin{equation}
    \begin{split}
        |\lambda_k\rangle=&\frac{1}{D\sqrt{p_k^\mu}}\sum_{j=0}^{D-1}e^{-i2\pi k j/D}|e^{i2\pi j/D} \sqrt{\mu}\rangle\\
        =&\frac{e^{-\frac{\mu}{2}}}{\sqrt{p_k^\mu}}\sum_{m=0}^\infty\frac{\sqrt{\mu}^{mD+k}}{\sqrt{(mD+k)!}}|mD+k\rangle,k=0,...,D-1,
    \end{split}
    \label{eq:ProtocolDescription_lambdak mu}
\end{equation}
and the probability distribution in the DPR case is given by
\begin{equation}
    \begin{split}
        p_k^\mu
        =&e^{-\mu}\sum_{m=0}^\infty\frac{\mu^{mD+k}}{(mD+k)!},
    \end{split}
    \label{eq:Protocol Description_pseudo Poisson distribution}
\end{equation}
which is a pseudo-Poisson distribution. \cref{eq:Source description_DPR channel} shows that the source output is diagonal in \(|\lambda_k\rangle\), \(k=0,...,D-1\), rather than number states. When \(D\) approaches infinity, the state \(|\lambda
_k\rangle\) in \cref{eq:ProtocolDescription_lambdak mu} approaches the photon-number state \(|k\rangle\), and the probability distribution in \cref{eq:Protocol Description_pseudo Poisson distribution} will also approach the Poisson distribution.
\par Next, we describe the two protocols considered in this paper, i.e., BB84 and MDI QKD using DPR.

\subsection{BB84 QKD with DPR \label{BB84 QKD with DPR}}
\par
\begin{figure}
    \centering
    \includegraphics[width=0.9\linewidth]{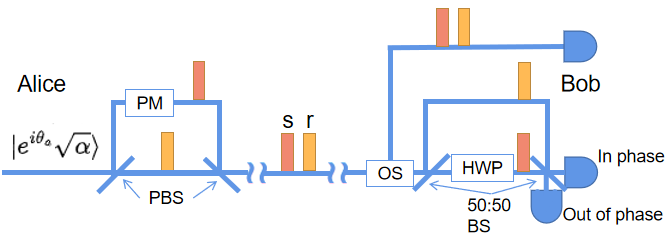}
    \caption{{Schematic diagram of the time-bin encoding BB84 protocol with DPR. Alice prepares a weak coherent pulse with a randomly chosen discrete phase \(\theta_a\in\{2\pi j/D, j=0,..., D-1\}\). By choosing the polarization of the input pulse, we can use a Mach-Zehnder interferometer to encode the input in $Z$ and $X$ bases. The resulting pulses are transmitted through a quantum channel to Bob. Upon receiving the pulses, Bob randomly selects his measurement basis. For the \(Z\) basis, he determines the logical bit by recording whether the detection event occurs in the early or late time bin. For the \(X\) basis, Bob measures the phase difference between r and s pulses. PBS: Polarizing beam splitter; PM: Phase modulator; OS: Optical switch; BS: Beam splitter; HWP: Half-wave plate.}}
    \label{fig:BB84 Scheme}
\end{figure}

 We study the BB84 with the vacuum+weak decoy state protocol using DPR. {We consider the time-bin encoding implementation of this protocol as shown in \cref{fig:BB84 Scheme}. In this scheme, the encoder output consists two pulses, denoted as reference (r) and signal (s). In $Z$ basis, a discrete phase randomized coherent state is sent in the reference (signal) pulse, with vacuum state in the other pulse, to represent bit 1 (0). In $X$ basis, a phase difference \(\{0,\pi\}\) is introduced between the signal and reference pulses. 

\cref{fig:BB84 Scheme} shows a schematic way to implement the above encoding and its corresponding decoding procedures. For each intensity \(\alpha\in\{0,\nu,\mu\}\), representing the vacuum, weak decoy, or the main signal, the input pulse to the encoder is uniformly chosen from the set of coherent states \(\{|e^{i2\pi j/D}\sqrt{\alpha}\rangle,j=0,...,D-1\}\). Encoding in $Z$ basis can be achieved by generating the input pulse in horizontal (vertical) polarization, and setting the phase modulator to zero. In $X$ basis, the input pulse to the encoder has diagonal polarization, therefore splitting at the polarizing beam splitter (PBS), and the phase modulator introduces a phase difference \(\{0,\pi\}\) between the signal and reference pulses.} {At the receiver, Bob either measures the time of arrival (for $Z$ basis) or uses a Mach-Zehnder interferometer to measure the phase difference between the reference and signal pulses ($X$ basis).} 

After the exchange of the raw key, Alice and Bob follow the typical steps in the vacuum+weak decoy state protocol to sift and reconcile their keys, and extract a secret key using privacy amplification. We only consider the asymptotic scenario in this work, which means that the intensity \(\mu\) is the main intensity to be sent and the probability of sending the weak decoy state or the vacuum state is negligibly small. We also use the efficient BB84 protocol \cite{razavi2018introduction}, where the \(Z\) basis is primarily used for secret key generation.
\par To model the outcome of the encoder in \cref{fig:BB84 Scheme}, we use the diagonal format of the phase randomized input state as shown in \cref{eq:Source description_DPR channel}. For a fixed intensity \(\mu\) and an encoded input state \(|\lambda_k\rangle\), with \(k=0,...,D-1\), the encoder output state in \(Z\) and \(X\) bases for bits \(0\) and \(1\) are, respectively, given by 
{
\begin{equation}
    \begin{split}
    |0_z\rangle_k&=\sum_{j=0}^{D-1}e^{-i2\pi k j/D}|\sqrt{\mu}e^{i2\pi j/D}\rangle_{s}|0\rangle_{r},\\
    |1_z\rangle_k&=\sum_{j=0}^{D-1}e^{-i2\pi k j/D}|0\rangle_{s}|\sqrt{\mu}e^{i2\pi j/D}\rangle_{r},\\
        |0_x\rangle_k&=\sum_{j=0}^{D-1}e^{-i2\pi k j/D}|\sqrt{\mu/2}e^{i2\pi j/D} \rangle_s|\sqrt{\mu/2}e^{i2\pi j/D} \rangle_r,\\
         |1_x\rangle_k&=\sum_{j=0}^{D-1}e^{-i2\pi k j/D}|\sqrt{\mu/2}e^{i(2\pi j/D+\pi)} \rangle_s|\sqrt{\mu/2}e^{i2\pi j/D} \rangle_r.
    \end{split}
    \label{eq:BB84 Prototocl_four emitted states with Z/X bases}
\end{equation}
Here,  \(|0_z\rangle_k\) and \(|1_z\rangle_k\)  (\(|0_x\rangle_k\) and \(|1_x\rangle_k\)) are logical qubits in the \(Z\) (\(X\)) basis for key bits \(0\) and \(1\), respectively, and for simplicity, we have not shown the normalization coefficients.}
\par For security purposes, it is essential that Eve cannot distinguish states in \(Z\) and \(X\) bases. In other words, the density operators for different bases should have an identical expression. This requirement is quantified by the fidelity between the corresponding density matrices \cite{cao2015discrete}:
\begin{equation}
    \begin{split}
                F_{k,\text{BB84}}=&\text{tr}\sqrt{\sqrt{\rho_{zk}}\rho_{xk}\sqrt{\rho_{zk}}},
                \label{eq:BB84 Source_mixed states of xy fidelity}
    \end{split}
\end{equation}where
\begin{align}
    \rho_{zk} &= |0_z\rangle_{kk}\langle0_z| + |1_z\rangle_{kk}\langle1_z|,\notag  \\
    \rho_{xk} &= |0_x\rangle_{kk}\langle0_x| + |1_x\rangle_{kk}\langle1_x|.
\end{align}
\par {Following the numerical methods in \cite{cao2015discrete}, we can verify that for \( k = 0 \) and \( k = 1 \), the above fidelity can, respectively, get close to \(1/\sqrt{2}\) and 1 when \(D\) increases.} This behavior is consistent with previous findings and supports the use of these components for key generation~\cite{lo2006security}. However, for \( k > 1 \), the fidelity significantly decreases (e.g., around \(0.35\) for \( k = 2 \) for parameter values used in \cite{cao2015discrete}), indicating stronger basis dependence. This implies that such states may not necessarily contribute much to the secure key rate. Therefore, in this work, we restrict key generation to the components with \( k = 0 \) and \( k = 1 \), where the basis dependence is often sufficiently low under practical conditions.

\subsection{MDI QKD with DPR \label{MDI QKD with DPR}}
\begin{figure}
    \centering
    \includegraphics[width=0.9\linewidth]{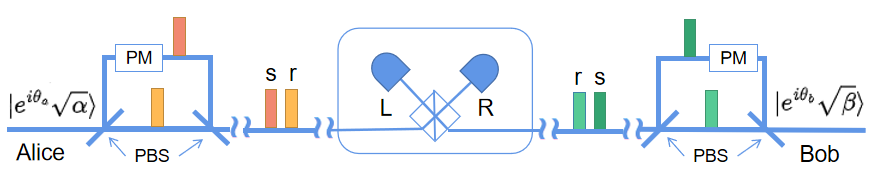}
    \caption{{Schematic diagram of the time-bin encoding MDI QKD protocol with DPR. Both Alice and Bob independently prepare weak coherent pulses with randomly chosen discrete phase values \(\theta_a ,\theta_b\in\{2\pi j/D,j=0,...,D-1\}\) with respect to their local phase references. Each prepared state is passed through a BB84 encoder as in \cref{fig:BB84 Scheme}. The resulting optical pulses are then transmitted through quantum channels to an untrustworthy intermediary, Charlie, who performs an interference measurement on the incoming states. Charlie publicly announces the detection results. Based on this announcement and their own encoding choices, Alice and Bob can post-select correlated events to extract a shared key bit.}}
    \label{fig:MDI Scheme}
\end{figure}
\par \cref{fig:MDI Scheme} shows the setup for the MDI QKD protocol considered in this paper. As in the case of BB84, we consider the time-bin encoding implementation of this protocol for the vacuum+weak decoy variant \cite{ma2012alternative}. Alice and Bob each uses a BB84 encoder as explained in \cref{BB84 QKD with DPR}, hence the output state from each encoder for a \(|\lambda_k\rangle\) input is given by \cref{eq:BB84 Prototocl_four emitted states with Z/X bases}. At the midpoint of the quantum channel, a third party, commonly referred to as Charlie, performs an interference measurement on the incoming reference and signal pulses by using single-photon detectors. A successful detection event is registered when only one of the two detectors clicks for both pairs of reference and signal pulses. Charlie then publicly announces the detection outcomes. Alice and Bob will decide on the shared bit based on the pattern of clicks reported by Charlie \cite{ma2012alternative}.  The post-processing steps are followed as described in \cref{BB84 QKD with DPR}.

{There are two additional points to mention for the particular implementation in \cref{fig:MDI Scheme}. First, note that there could be a random offset between the phase reference at Alice versus that of Bob. The phase values in \cref{eq:BB84 Prototocl_four emitted states with Z/X bases} are therefore with respect to a local phase reference. Second, given that the scheme in \cref{fig:MDI Scheme} relies on interference, polarization maintenance needs to be employed to ensure maximum interference at the middle beam splitter. We assume that this is the case in our analysis.}

\par To ensure that the bases are mutually unbiased the density operators in the \(Z\) and \(X\) bases must be identical. This requirement is quantified by the fidelity between the corresponding density matrices \cite{cao2020discrete} as follows:
\begin{equation}
    \begin{split}
        F_{k,\text{MDI}}=\text{tr}\sqrt{\sqrt{\rho_{zk}}\rho_{xk}\sqrt{\rho_{zk}}},
    \end{split}
    \label{eq:MDI Source_mixed states of xy fidelity}
\end{equation}where
\begin{equation}
    \begin{split}
         &\rho_{zk}\\=&(|0_{z}\rangle_{kk}\langle0_{z}|+|1_{z}\rangle_{kk}\langle1_{z}|)_A\otimes(|0_{z}\rangle_{kk}\langle0_{z}|+|1_{z}\rangle_{kk}\langle1_{z}|)_B,\\
        &\rho_{yk}\\=&(|0_{x}\rangle_{kk}\langle0_{x}|+|1_{x}\rangle_{kk}\langle1_{x}|)_A\otimes(|0_{x}\rangle_{kk}\langle0_{x}|+|1_{x}\rangle_{kk}\langle1_{x}|)_B.
    \end{split}
\end{equation}

\begin{mybox}[Box:NonlinearOptimizationBB84]{Nonlinear optimization problem for DPR BB84 protocol with vacuum+weak decoy states}
\begin{center}
    \textbf{Generic Formulation}
\end{center}
\textbf{Minimize}
\begin{align}
        R^\text{DPR}_\text{BB84}=\sum_{k=\{0,1\}} Q_{z,k}^\mu[1 - h_2(e_{z,k}^{p,\mu})] - f Q_z^\mu h_2(E_z^\mu).\notag
\end{align}
\textbf{subject to}
\begin{align}
         &Q_\gamma^{\alpha} = \sum_{k=0}^{D-1} p_k^{\alpha} Y_{\gamma,k}^{\alpha},\notag \\& Q_\gamma^\alpha E_\gamma^{\alpha} = \sum_{k=0}^{D-1} p_k^{\alpha} e_{\gamma,k}^{b,\alpha} Y_{\gamma,k}^{\alpha},\notag\\
        &|Y_{\gamma,k}^{\alpha_1} - Y_{\gamma,k}^{\alpha_2}| \leq \epsilon_{\alpha_1\alpha_2}, \quad{k=0,\ldots, D-1} \notag\\
        &|e_{\gamma,k}^{b,\alpha_1} Y_{\gamma,k}^{\alpha_1} - e_{\gamma,k}^{b,\alpha_2} Y_{\gamma,k}^{\alpha_2}| \leq \epsilon_{\alpha_1\alpha_2}, \quad{k=0,\ldots, D-1}\notag\\
      &e^{p,\mu}_{z,k} \leq e^{b,\mu}_{x,k} + 4\Delta^{\mu}_{k}(1 - \Delta^{\mu}_{k})(1 - 2e^{b,\mu}_{x,k}) \notag\\
        &\quad+ 4(1 - 2\Delta^{\mu}_{k}) \sqrt{\Delta^{\mu}_{k}(1 - \Delta^{\mu}_{k}) e^{b,\mu}_{x,k}(1 - e^{b,\mu}_{x,k})}, k=0,1\notag
    \end{align}
where \(\alpha,\alpha_1,\alpha_2\in\{0,\nu,\mu\}\), \(\gamma=\{z,x\}\) represents \(Z\) or \(X\) basis, \(p_k^\mu\) is given in \cref{eq:Protocol Description_pseudo Poisson distribution}, \(\epsilon_{\alpha_1\alpha_2}\) comes from \cref{eq:BB84 Parameter_bound of y or eb with different intensities}, and \(\Delta_k^{\mu}\) is defined in \cref{eq:KeyRateAnalysis_Deltaprime}.
\begin{center}
\textbf{Approximate Formulation}
\end{center}
As above, with the additional constraint that \(Y_{\gamma,k}^\alpha=0\), for \(k=3,4,...,D-1\).
\end{mybox}

\section{Key Rate Analysis \label{Key rate analysis}}
\par In this section, we review the key rate analysis for the two protocols described in \cref{Protocol description}. The corresponding optimization problems are summarized in \cref{Box:NonlinearOptimizationBB84} and \cref{Box:NonlinearOptimizationMDI}, respectively, for BB84 and MDI QKD protocols. Our analytical solutions to these problems are then given in \cref{Analytical estimation}.

\subsection{BB84 with DPR: Rate analysis \label{BB84 QKD key rate analysis}}
\par The key rate of the BB84 protocol in \cref{BB84 QKD with DPR} is lower bounded by \cite{gottesman2004security}
\begin{align}
    R^\text{DPR}_\text{BB84} &= \sum_{k=\{0,1\}} Q_{z,k}^\mu[1 - h_2(e_{z,k}^{p,\mu})] - f Q_z^\mu h_2(E_z^\mu),
    \label{eq:BB84 Protocol_RDBB84}
\end{align}
where \(Q_{z,k}^\mu\) denotes the gain when Alice encodes the state \(|\lambda_k\rangle\) with intensity \(\mu\), \(f\) is the error correction inefficiency, and \(h_2(x)= -x \log_2(x) - (1 - x) \log_2(1 - x)\) is the Shannon binary entropy function. The term \(e_{z,k}^{p,\mu}\) denotes the phase error rate for \(|\lambda_k\rangle\) with intensity \(\mu\) as the input state, which in the \(Z\) basis corresponds to errors caused by Pauli operator \(\sigma_z\). The second term in the key rate expression accounts for the cost of error correction, where \(Q_z^\mu\) and \(E_z^\mu\) are the total gain and quantum bit error rate (QBER), respectively, for intensity \(\mu\). All above terms are for the \(Z\) basis, which is the main basis used in the efficient QKD protocol we defined in \cref{BB84 QKD with DPR}.
\par To compute the key rate, two critical parameters must be bounded: The yield term \(Y_{z,k}^\mu = Q_{z,k}^\mu/p_k^\mu\), and the phase error rate \(e^{p,\mu}_{z,k}\). Both parameters are associated with the quantum state \(|\lambda_k\rangle\), rather than the photon number state \(|k\rangle\) in CPR. Another distinct feature in DPR is that, unlike the CPR case, the yield, gain, and error parameters corresponding to \(|\lambda_k\rangle\) are now functions of the employed intensity as well. This will make the optimization problem more tedious to solve as we see below. 
\par To minimize the lower bound in \cref{eq:BB84 Protocol_RDBB84}, we subject it to the following constraints set by the experimental observables:
\begin{align}
    Q_\gamma^{0} &= Y_{\gamma,0}^{0},  & Q_\gamma^0E_\gamma^{0} &= e_{\gamma,0}^{b,0}Y_{\gamma,0}^{0}; \notag \\
    Q_\gamma^{\nu} &= \sum_{k=0}^{D-1} p_k^{\nu}Y_{\gamma,k}^{\nu},  & Q_\gamma^\nu E_\gamma^{\nu} &= \sum_{k=0}^{D-1} p_k^{\nu}e_{\gamma,k}^{b,\nu}Y_{\gamma,k}^{\nu}; \notag \\
    Q_\gamma^{\mu} &= \sum_{k=0}^{D-1} p_k^{\mu}Y_{\gamma,k}^{\mu},  & Q_\gamma^\mu E_\gamma^{\mu} &= \sum_{k=0}^{D-1} p_k^{\mu}e_{\gamma,k}^{b,\mu}Y_{\gamma,k}^{\mu},
    \label{eq:ParameterDifferenceinCPRandDPR_DPR nonlinear problem}
\end{align}
where the subscript \(\gamma=\{z,x\}\) specfies the basis and \(e_{\gamma,k}^{b,\alpha}\) is the bit error rate when \(|\lambda_k\rangle\), with intensity \(\alpha\), is at the encoder input.
\par To minimize \cref{eq:BB84 Protocol_RDBB84} subject to \cref{eq:ParameterDifferenceinCPRandDPR_DPR nonlinear problem}, we use the following inequalities \cite{gottesman2004security, cao2015discrete}:
\begin{align}
    |Y_{\gamma,k}^{\alpha_1}-Y_{\gamma,k}^{\alpha_2}| &\leq \sqrt{1-F_{\alpha_1\alpha_2}^2} \equiv \epsilon_{\alpha_1\alpha_2}, \notag \\
    |e_{\gamma,k}^{b,\alpha_1}Y_{\gamma,k}^{\alpha_1}-e_{\gamma,k}^{b,\alpha_2}Y_{\gamma,k}^{\alpha_2}| &\leq \sqrt{1-F_{\alpha_1\alpha_2}^2} \equiv \epsilon_{\alpha_1\alpha_2}.
    \label{eq:BB84 Parameter_bound of y or eb with different intensities}
\end{align}
where \(\alpha_1\) and \(\alpha_2\) are two intensity variables in \(\{0,\nu,\mu\}\), and the fidelity \(F_{\alpha_1\alpha_2}\) is given by
\begin{align}
    F_{\alpha_1\alpha_2} = \frac{\sum_{m=0}^\infty \frac{(\alpha_1\alpha_2)^{mD/2}}{(mD)!}}{\sqrt{\sum_{m_1=0}^\infty \frac{\alpha_1^{m_1D}}{(m_1D)!} \sum_{m_2=0}^\infty \frac{\alpha_2^{m_2D}}{(m_2D)!}}}.
    \label{eq:BB84 key rate anlaysis_fidelity between alpha1 and alpha2}
\end{align}
\par Finally, to bound the privacy amplification term, the phase error rate term \(e^{p,\mu}_{z,k}\) needs to be upper bounded. We use the virtual protocol described in \cite{gottesman2004security,koashi2005simple} to establish a relationship between the bit error rate \(e^{b,\mu}_{x,k}\) in the \(X\) basis and the phase error rate \(e^{p,\mu}_{z,k}\) in the \(Z\) basis, which characterizes the effective basis dependence in realistic scenarios by the following parameter:
\begin{equation}
    \begin{split}
        \Delta_k^{\mu}=\text{min}(1,\frac{1-F_{k,\text{BB84}}}{2Y_{z,k}^\mu}).
    \end{split}
    \label{eq:KeyRateAnalysis_Deltaprime}
\end{equation}
Using a bound proved in \cite{lo2006security}, we can get a relationship between \(e^{b,\mu}_{x,k}\) and \(e^{p,\mu}_{z,k}\) as follows:
\begin{align}
    e^{p,\mu}_{z,k} &\leq e^{b,\mu}_{x,k} + 4\Delta^{\mu}_{k}(1 - \Delta^{\mu}_{k})(1 - 2e^{b,\mu}_{x,k}) + \notag \\
    &  4(1 - 2\Delta^{\mu}_{k}) \sqrt{\Delta^{\mu}_{k}(1 - \Delta^{\mu}_{k}) e^{b,\mu}_{x,k}(1 - e^{b,\mu}_{x,k})}, k=0,1.
    \label{eq:BB84_bound between ep1 and eb1}
\end{align}
\par Putting everything together, in \cref{Box:NonlinearOptimizationBB84}, we have summarized the generic formulation of the optimization problem that we have to solve. Numerically solving this problem can be cumbersome as \(Y_{\gamma,k}^\alpha\) terms can be non-negative for all values of \(k=0,...,D-1\). An approximation to this problem can be achieved if we assume \(Y_k^\alpha=0\) for \(k\geq3\). This is how we find the numerical bounds on the key rate in \cref{Simulation}. In \cref{Analytical estimation}, however, we derive analytical bounds on the relevant terms in the generic formulation of the key rate optimization problem. It not only reduces the required computational resources, but also its results closely match the results obtained by numerical optimization in the regions of interest.

\vspace{1em}
\begin{mybox}[Box:NonlinearOptimizationMDI]{Nonlinear optimization problem for DPR MDI QKD with vacuum+weak decoy states}
\begin{center}
    \textbf{Generic Formulation}
\end{center}
\textbf{Minimize}
\begin{align}
    R^\text{DPR}_\text{MDI}= \sum_{k=\{0,1\}} Q_{z,kk}^{\mu\mu}[1 - h_2(e_{z,kk}^{p,\mu\mu})] - f Q_z^{\mu\mu} h_2(E_z^{\mu\mu}).\notag
\end{align}
\textbf{subject to}
\begin{align}
       & Q_\gamma^{\alpha\beta} = \sum_{k,l=0}^{D-1} p_k^\alpha p_l^\beta Y_{\gamma,kl}^{\alpha\beta},\notag\\
       &Q_\gamma^{\alpha\beta}E_\gamma^{\alpha\beta} = \sum_{k,l=0}^{D-1} p_k^\alpha p_l^\beta e_{\gamma,kl}^{b,\alpha\beta} Y_{\gamma,kl}^{\alpha\beta},\notag\\
    &|e_{\gamma,kl}^{b,\alpha_1\beta_1} Y_{\gamma,kl}^{\alpha_1\beta_1} - e_{\gamma,kl}^{b,\alpha_2\beta_2} Y_{\gamma,kl}^{\alpha_2\beta_2}| \leq \epsilon_{\alpha_1\alpha_2} + \epsilon_{\beta_1\beta_2},\notag\\
    &|Y_{\gamma,kl}^{\alpha_1\beta_1} - Y_{\gamma,kl}^{\alpha_2\beta_2}| \leq \epsilon_{\alpha_1\alpha_2} + \epsilon_{\beta_1\beta_2}, \quad{k,l=0,\ldots, D-1}\notag\\
         &e^{p,\mu\mu}_{z,kk} \leq e^{b,\mu\mu}_{x,kk} + 4\Delta^{\mu\mu}_{kk}(1 - \Delta^{\mu\mu}_{kk})(1 - 2e^{b,\mu\mu}_{x,kk}) +\notag\\
        & 4(1 - 2\Delta^{\mu\mu}_{kk}) \sqrt{\Delta^{\mu\mu}_{kk}(1 - \Delta^{\mu\mu}_{kk}) e^{b,\mu\mu}_{x,kk}(1 - e^{b,\mu\mu}_{x,kk})},k=0,1\notag
    \end{align}
where \(\alpha,\alpha_1,\alpha_2,\beta,\beta_1,\beta_2\in\{0,\nu,\mu\}\), \(\gamma=\{z,x\}\) represents either \(Z\) or \(X\) basis, \(p_k^\mu\), \(p_l^\beta\) are given in \cref{eq:Protocol Description_pseudo Poisson distribution}, \(\epsilon_{\alpha_1\alpha_2}\) and \(\epsilon_{\beta_1\beta_2}\) come from \cref{eq:BB84 Parameter_bound of y or eb with different intensities}, and \(\Delta^{\mu\mu}_{kk}\) is defined in \cref{eq:KeyRateAnalysis_Deltaprimeab}.
\begin{center}
    \textbf{Approximate Formulation}
\end{center}
As above, with the additional constraint that \(Y_{\gamma,kl}^{\alpha\beta}=0\), for \(k,l=7,8,...,D-1\).
\end{mybox}

\subsection{MDI QKD with DPR: Rate analysis \label{MDI QKD key rate analysis}}
\par Considering the DPR scenario, the secret key rate for the MDI QKD protocol is lower bounded by \cite{gottesman2004security, cao2020discrete}:
\begin{equation}
        R^\text{DPR}_\text{MDI}=\sum_{k=\{0,1\}} Q_{z,kk}^{\mu\mu}[1-h_2(e_{z,kk}^{p,\mu\mu})]-fQ_z^{\mu\mu} h_2(E_z^{\mu\mu}),
    \label{eq:MDI Protocol_RDMDI}
\end{equation}
where, in general, \(Q_{z,kl}^{\mu\mu}\) and \(e_{z,kl}^{p,\mu\mu}\), respectively, are the gain and phase error rate, when Alice and Bob encode \(|\lambda_k\rangle\) and \(|\lambda_l\rangle\), both with intensity $\mu$, in the \(Z\) basis, and \(Q_z^{\mu\mu}\) and \(E_z^{\mu\mu}\) are, respectively, the total gain and QBER when intensity \(\mu\) is used by both Alice and Bob in the \(X\) basis. 
{As in the case of BB84, in \cref{eq:MDI Protocol_RDMDI}, we only include terms that correspond to $|\lambda_0\rangle$ and $|\lambda_1\rangle$. In particular, we have only accounted for the terms that correspond to sending the same state by both Alice and Bob. We can numerically verify that all other terms contribute negligibly to the key rate in regions of interest. The choice of \cref{eq:MDI Protocol_RDMDI}, as the lower bound, also simplifies deriving analytical bounds for the relevant terms.}

\par First, we note that the following relations hold:
\begin{equation}
    \begin{split}
          Q_\gamma^{\alpha\beta}=\sum_{k,l=0}^{D-1}p_k^\alpha p_l^\beta Y_{\gamma,kl}^{\alpha\beta}, 
        Q_\gamma^{\alpha\beta}E_\gamma^{\alpha\beta}=\sum_{k,l=0}^{D-1}p_k^\alpha p_l^\beta e_{\gamma,kl}^{b,\alpha\beta}Y_{\gamma,kl}^{\alpha\beta},
    \end{split}
    \label{eq:MDI key rate analysis_DPR nonlinear equations}
\end{equation}
where \(\gamma=\{z,x\}\) represents \(Z\) or \(X\) basis, and \(Q_\gamma^{\alpha\beta}\) and \(E_\gamma^{\alpha\beta}\) are experimentally observable variables in basis $\gamma$, respectively, representing the total gain and QBER, when Alice and Bob use intensities \(\alpha\) and \(\beta\). The term \(Y_{\gamma,kl}^{\alpha\beta}\) denotes the yield in basis $\gamma$ when Alice and Bob encode the states \(|\lambda_k\rangle\), with intensity $\alpha$, and \(|\lambda_l\rangle\), with intensity $\beta$, respectively, and \(e_{\gamma,kl}^{b,\alpha\beta}\) is the corresponding bit error rate. 
\par Similar to \cref{eq:BB84 Parameter_bound of y or eb with different intensities}, in the MDI case, we have the following relationships \cite{cao2015discrete}:
\begin{align}
    |Y_{\gamma,kl}^{\alpha_1\beta_1} - Y_{\gamma,kl}^{\alpha_2\beta_2}| &\leq \epsilon_{\alpha_1\alpha_2} + \epsilon_{\beta_1\beta_2}, \notag \\
    |e_{\gamma,kl}^{b,\alpha_1\beta_1}Y_{\gamma,kl}^{\alpha_1\beta_1} - e_{\gamma,kl}^{b,\alpha_2\beta_2}Y_{\gamma,kl}^{\beta_2}| &\leq \epsilon_{\alpha_1\alpha_2} + \epsilon_{\beta_1\beta_2}.
    \label{eq:MDI Parameter_bound of y or eb with different intensities}
\end{align}
\par The final step is to bound the phase error rate \(e^{p,\mu\mu}_{z,kk}\) in the \(Z\) basis by the bit error rate \(e_{x,kk}^{b,\mu\mu}\) in the \(X\) basis. Similar to what we did in \cref{BB84 QKD key rate analysis}, the basis dependence can be characterized by \cite{gottesman2004security,koashi2005simple}:
\begin{equation}
    \begin{split}
     \Delta_{kk}^{\mu\mu}=\text{min}(1,\frac{1-F_{k,\text{MDI}}}{2Y_{z,kk}^{\mu\mu}}),
    \end{split}
    \label{eq:KeyRateAnalysis_Deltaprimeab}
\end{equation}
which corresponds to the case when both Alice and Bob encode \(|\lambda_k\rangle\). Using the above, it can be shown that  \cite{tamaki2003unconditionally,lo2006security,cao2020discrete}:
\begin{equation}
    \begin{split}
        &e^{p,\mu\mu}_{z,kk}\leq e^{b,\mu\mu}_{x,kk}+4\Delta^{\mu\mu}_{kk}(1-\Delta^{\mu\mu}_{kk})(1-2e^{b,\mu\mu}_{x,kk}) \\ +
        &4(1-2\Delta^{\mu\mu}_{kk})\sqrt{\Delta^{\mu\mu}_{kk}(1-\Delta^{\mu\mu}_{kk})e^{b,\mu\mu}_{x,kk}(1-e^{b,\mu\mu}_{x,kk})}, k=0,1.
    \end{split}
    \label{eq:MDI_bound between ep11 and eb11}
\end{equation}
\par Finally, we can construct the numerical optimization problem in \cref{Box:NonlinearOptimizationMDI} to work out the key rate of DPR MDI QKD protocol. The approximate formulation used for numerically solving this problem is also shown in \cref{Box:NonlinearOptimizationMDI}. The analytical methods are discussed in \cref{Analytical MDI} and \cref{Appendix: analtycial MDI}.

\section{Analytical Bounds \label{Analytical estimation}}
\par In this section, we present analytical bounds on the key parameters in \cref{eq:BB84 Protocol_RDBB84} and \cref{eq:MDI Protocol_RDMDI}. The key objective is to derive bounds on such parameters using experimentally accessible observables from vacuum, decoy, and signal states. Our approach parallels the methodology used in CPR QKD \cite{ma2005practical,gobby2004quantum,sun2013practical}, but requires additional care due to the differences introduced by DPR \cite{cao2020discrete}. Here, we summarize the final results leaving detailed derivations to \cref{Appendix: analytical BB84} and \cref{Appendix: analtycial MDI}.

\subsection{Analytical bounds for BB84 with DPR \label{Analytical BB84}}
\par In this section, we give analytical bounds on $Y_{\gamma,0}^\mu$, $Y_{\gamma,1}^\mu$, $W_{x,0}^\mu$, and $W_{x,1}^\mu$, with $W_{\gamma,k}^\alpha = e_{\gamma,k}^{b,\alpha} Y_{\gamma,k}^{\alpha}$, which contribute to the key rate expression in \cref{eq:BB84 Protocol_RDBB84}.
\par First, note that \(Q_\gamma^0=Y_{\gamma,0}^0\) and \(Q_\gamma^0E_\gamma^0=W_{\gamma,0}^0\). Using \cref{eq:BB84 Parameter_bound of y or eb with different intensities}, we have
\begin{align}
    Y_{\gamma,0}^\mu &\geq Y_{\gamma,0}^{\mu L} \equiv \max(Q_\gamma^0 - \epsilon_{\mu 0}, 0), \notag \\
    W_{yx,0}^\mu &\leq W_{x,0}^{\mu U} \equiv \min(Q_x^0E_x^0 + \epsilon_{\mu 0}, 0.5),
    \label{eq:Analytical BB84_Y0 and eb0}
\end{align}
where  \(Q_\gamma^0\) and \(Q_x^0E_x^0\) are directly obtained from observables. Using these, we can derive an upper bound for the bit error rate as \(e_{x,0}^{b,\mu}\leq e_{x,0}^{b,\mu U} \equiv W_{x,0}^{\mu U}/Y_{x,0}^{\mu L}\).
\par The next key parameters in the analysis are \(Y_{\gamma,1}^\mu\) and \(W_{x,1}^\mu = e_{x,1}^{b,\mu} Y_{x,1}^\mu\), which correspond to the state \(|\lambda_1\rangle\). The lower bound on \(Y_{\gamma,1}^\mu\) is given by:
\begin{align}
        &Y_{\gamma,1}^\mu\geq Y_{\gamma,1}^{\mu L}\notag\\\equiv&\frac{e^{\alpha_1} Q_\gamma^{\alpha_1}-e^{\alpha_2} Q_\gamma^{\alpha_2}-(e^{\alpha_1} p_0^{\alpha_1}-e^{\alpha_2}p_0^{\alpha_2}-A_{\alpha_1\alpha_2}e^\mu p_0^\mu)Q_\gamma^0}{(e^{\alpha_1} p_1^{\alpha_1}-e^{\alpha_2}p_1^{\alpha_2}-A_{\alpha_1\alpha_2}e^\mu p_1^\mu)}\notag\\&-\frac{A_{\alpha_1\alpha_2}e^{\mu}Q^\mu + A_{\alpha_1\alpha_2}e^\mu p_0^\mu \epsilon_{\mu 0}+e^{\alpha_1}p_0^{\alpha_1}\epsilon_{\alpha_1 0}+e^{\alpha_2}p_0^{\alpha_2}\epsilon_{\alpha_2 0}}{(e^{\alpha_1} p_1^{\alpha_1}-e^{\alpha_2}p_1^{\alpha_2}-A_{\alpha_1\alpha_2}e^\mu p_1^\mu)}\notag\\&-\frac{(e^{\alpha_1}-e^{\alpha_1}p_0^{\alpha_1})\epsilon_{\mu\alpha_1}+(e^{\alpha_2}-e^{\alpha_2}p_0^{\alpha_2})\epsilon_{\mu\alpha_2}}{(e^{\alpha_1} p_1^{\alpha_1}-e^{\alpha_2}p_1^{\alpha_2}-A_{\alpha_1\alpha_2}e^\mu p_1^\mu)},
    \label{eq:Analytical BB84_Lower bound of Y1mu}
\end{align}
where
\begin{equation}
    \begin{split}
        A_{\alpha_1\alpha_2} = \max_{k \geq 2} \left\{ \frac{e^{\alpha_1}p_k^{\alpha_1} - e^{\alpha_2}p_k^{\alpha_2}}{e^{\mu}p_k^{\mu}} \right\},
    \end{split}
     \label{eq:Appendix_BB84_Parameter_A}
\end{equation}
and we set \(\alpha_1=\nu\) and \(\alpha_2=0\), because they often give the best results. In \cref{eq:Appendix_BB84_Parameter_A}, we can prove that $A_{\alpha_1\alpha_2}$ is monotonically decreasing with \(k\) when \(\alpha_1>\alpha_2\) and achieves its maximal value at \(k=2\). The above formulation nevertheless gives us the option to explore alternative combinations of $\alpha_1$ and $\alpha_2$ without compromising the security of the protocol.

\par {Next, the upper bound on \(W_{x,1}^\mu\) is given by 
\begin{equation}
    \begin{split}
        &W_{x,1}^\mu\leq W_{x,1}^{\mu U}\\ &\equiv\frac{Q_x^{\alpha_1}E_x^{\alpha_1}-p_0^{\alpha_1}Q_{x,0}^0E_{x,0}^0+p_0^{\alpha_1}\epsilon_{\alpha_1 0}+p_1^{\alpha_1}\epsilon_{\mu\alpha_1}}{p_1^{\alpha_1}}.
    \end{split}
    \label{eq:Analytical BB84_Upper bound of W1mu}
\end{equation}
Consequently, the upper bound on the bit error rate is given by \(e^{b,\mu U}_{x,1}=W_{x,1}^{\mu U}/Y_{x,1}^{\mu L}\).}

\par In summary, by employing the vacuum+weak decoy state method and applying fidelity-based constraints, we are able to derive lower bounds for the yields \(Y_{\gamma,0}^\mu\) and \(Y_{\gamma,1}^\mu\), as shown in \cref{eq:Analytical BB84_Y0 and eb0} and \cref{eq:Analytical BB84_Lower bound of Y1mu}, respectively. Similarly, upper bounds for the bit error rates \(e_{x,0}^{b,\mu}\) and \(e_{x,1}^{b,\mu}\) are obtained by using \cref{eq:Analytical BB84_Y0 and eb0} and \cref{eq:Analytical BB84_Upper bound of W1mu}. With these bounds, the lower bound on the key rate can be calculated by using \cref{eq:BB84 Protocol_RDBB84}, where the phase error rate of the \(Z\) basis is bounded via the bit error rate in the \(X\) basis, as described in \cref{eq:BB84_bound between ep1 and eb1}. All other required parameters in \cref{eq:BB84 Protocol_RDBB84} are either directly measurable in experiments or computable through simulations.

\subsection{Analytical bounds for MDI QKD with DPR \label{Analytical MDI}}
\par In the DPR case, a lower bound on the key rate of the MDI-QKD protocol is given by \cref{eq:MDI Protocol_RDMDI}. The variables we need to bound are \(Y_{\gamma,kk}^{\mu\mu}=Q_{\gamma,kk}^{\mu\mu}/(p_k^\mu p_k^\mu)\) and \(e_{x,kk}^{b,\mu\mu}\), for \(k\in\{0,1\}\), from which \(e_{z,kk}^{p,\mu\mu}\) can be bounded.
\par The bounds for variables related to \(|\lambda_{00}\rangle\) are expressed as follows:
\begin{align}
    Y_{\gamma,00}^{\mu\mu} &\geq Y_{\gamma,00}^{\mu\mu L} \equiv \max(Q_\gamma^{00} - 2\epsilon_{\mu0}, 0), \notag \\
    W_{x,00}^{\mu\mu} &\leq W_{x,00}^{\mu\mu U} \equiv \max(Q_x^{00}E_x^{00} + 2\epsilon_{\mu0}, 0.5),
    \label{eq:Analytical MDI bounds_Y00 and W00}
\end{align}
where \(Q_x^{00}\) and \(Q_x^{00}E_x^{00}\) are observables, and \(W_{\gamma,k}^{\alpha\beta}=e_{\gamma,k}^{b,\alpha\beta} Y_{\gamma,k}^{\alpha\beta}\). An upper bound for the bit error rate is \(e^{b,\mu\mu}_{x,00}\leq e^{b,\mu\mu U}_{x,00}= W_{x,00}^{\mu\mu U}/Y_{x,00}^{\mu\mu L}\).
\par In terms of bounds on \(Y_{\gamma,11}^{\mu\mu}\) and \(W_{x,11}^{\mu\mu}\), we use \cref{eq:MDI Parameter_bound of y or eb with different intensities} and some mathematical inequalities in \cref{Appendix: analtycial MDI} to obtain
\begin{equation}
    \begin{split}
         Y_{\gamma,11}^{\mu\mu}\geq Y_{\gamma,11}^{\mu\mu L}\equiv \frac{e^{\alpha+\beta} Q_\gamma^{\alpha\beta}-T_1-GT_s-\epsilon_{\bar{s}}-2G\epsilon_{T_s}-\epsilon_{s}}{e^{\alpha+\beta}p_1^{\alpha}p_1^{\beta}-G(e^{\mu}p_1^{\mu})^2},
    \end{split}
    \label{eq:Analytical MDI_Y11L}
\end{equation}
where 
\begin{align}
    T_1 &= e^{\alpha+\beta}(p_0^\alpha Q_\gamma^{0\beta} + p_0^\beta Q_\gamma^{\alpha 0} - p_0^\alpha p_0^\beta Q_\gamma^{00})\quad\text{and} \notag \\
    T_s &= e^{2\mu}(Q_\gamma^{\mu\mu} - p_0^\mu(Q_\gamma^{0\mu} + Q_\gamma^{\mu 0}) + p_0^{\mu}p_0^{\mu}Q_\gamma^{00})
    \label{eq:Analytical MDI_Discrete ealbe Qalbe}
\end{align}
can be calculated from observed values; \(G=\text{max}(A,B,C)\), where
\begin{align}
    A &= \max_{k \geq 2} \left\{ \frac{e^{\alpha+\beta}p_k^\alpha p_1^\beta}{e^{2\mu}p_k^\mu p_1^\mu} \right\}, \quad
    B = \max_{l \geq 2} \left\{ \frac{e^{\alpha+\beta}p_1^\alpha p_l^\beta}{e^{2\mu}p_1^\mu p_l^\mu} \right\}, \notag \\
    C &= \max_{k,l \geq 2} \left\{ \frac{e^{\alpha+\beta}p_k^\alpha p_l^\beta}{e^{2\mu}p_k^\mu p_l^\mu} \right\}.
    \label{eq:Analytical MDI_Discrete bounds A B C}
\end{align}
Here, $A$, $B$, and \(C\) are monotonically decreasing with respect to $k$ and $l$ and achieve their maximal values at \(k=l=2\). Also, from \cref{eq:MDI Parameter_bound of y or eb with different intensities}, we show in \cref{Appendix: analtycial MDI} that 
\begin{align}
        \epsilon_{\bar{s}}=&e^{\alpha+\beta}(p_0^{\alpha}+p_0^{\alpha}p_0^{\beta})\epsilon_{\alpha 0}+e^{\alpha+\beta}(p_0^{\beta}+p_0^{\alpha}p_0^{\beta})\epsilon_{\beta 0}\notag\\&+e^{\alpha+\beta}p_1^{\alpha}p_1^{\beta}(\epsilon_{\alpha\mu}+\epsilon_{\beta\mu}),\notag\\
        \epsilon_{s}=&e^{\alpha+\beta}(\sum_{a=2}^{D-1}p_a^\alpha p_1^\beta(\epsilon_{\alpha\mu}+\epsilon_{\beta\mu})+\sum_{b=2}^{D-1}p_1^\alpha p_b^\beta (\epsilon_{\alpha\mu}+\epsilon_{\beta\mu})\notag\\&+\sum_{a=2}^{D-1}\sum_{b=2}^{D-1}p_a^\alpha p_b^\beta (\epsilon_{\alpha\mu}+\epsilon_{\beta\mu})),\notag\\
        \epsilon_{T_s}=&e^{2\mu}(p_0^{\mu}+p_0^{\mu}p_0^{\mu})\epsilon_{\mu 0}.
        \label{eq:MDI Analytical Bound_epsilons}
    \end{align}
One important condition is \(e^{\alpha+\beta}p_1^{\alpha}p_1^{\beta}-G(e^{\mu}p_1^{\mu})^2>0\) to let the value of \(Y_{\gamma,11}^{\mu\mu L}\) be positive and meaningful. After calculation, we can prove that as long as \(\alpha\) and \(\beta\) are not both equal to \(\mu\), this condition holds, which is the reason why decoy states must be involved in the analytical methods. As a result, in this work, we use \(\alpha=\beta=\nu\) to analyze the key rate.
\par Next, the upper bound on \(W_{x,11}^{\mu\mu}\) is calculated from
\begin{equation}
    \begin{split}
         W_{x,11}^{\mu\mu}\leq W_{x,11}^{\mu\mu U}\equiv\frac{e^{\alpha+\beta} Q_x^{\alpha\beta}E_x^{\alpha\beta}-T_2+\epsilon_{\bar{s}}}{e^{\alpha+\beta}p_1^{\alpha}p_1^{\beta}},
    \end{split}
    \label{eq:Analytical MDI bounds_W11U}
\end{equation}
where
\begin{equation}
           T_2=e^{\alpha+\beta}(p_0^{\alpha}Q_x^{0\beta}E_x^{0\beta}+p_0^{\beta}  Q_x^{\alpha 0}E_x^{\alpha 0}-p_0^{\alpha}p_0^{\beta}Q_x^{00}E_x^{00}).
       \label{eq:MDI Analytical Bound_T2}
\end{equation}
 \section{Simulation Results and Discussion \label{Simulation}} 
\par In our simulations, we consider the scenario in which no eavesdropper (Eve) is present, but system parameters and imperfections determine the observed values in an experiment. In our simulations of the BB84 protocol, the total gain and the corresponding QBER, for an intensity \(\alpha\), are, respectively, assumed to be~\cite{panayi2014memory}
\begin{align}
    Q_\gamma^\alpha &=1 - e^{-\eta\alpha}(1-p_d)^2, \notag \\
    Q_\gamma^\alpha E_\gamma^\alpha &= e_0 Q_\gamma^\alpha - (e_0 - e_d)(1 - e^{-\eta\alpha})(1 - p_d),
    \label{eq:Simulation BB84_Q QE}
\end{align}
where \(p_d\) is the dark count rate per pulse, \(\eta\) represents the overall channel transmittance, \(e_0 = 0.5\) is the error rate associated with vacuum states, and \(e_d\) denotes the phase misalignment error rate. The nominal values for the simulation parameters are shown in \cref{table:system_parameters} \cite{gobby2004quantum,cao2015discrete}, where the optical switch insertion loss is included in the detector efficiency.

In our simulations of the MDI QKD protocol, the total gain and the corresponding QBER when Alice sends states with intensity \(\alpha\) and Bob sends states with intensity \(\beta\) are, respectively, taken {as~\cite{ma2012statistical}
\begin{align}
        Q_z^{\alpha\beta}=&Q_C+Q_E,\notag\\
        Q_z^{\alpha\beta} E_z^{\alpha\beta}=&e_dQ_C+(1-e_d)Q_E,\notag\\
        Q_x^{\alpha\beta}=&2y^2[1+2y^2-4yI_0(x)+I_0(2x)],\notag\\
         Q_x^{\alpha\beta} E_x^{\alpha\beta}=&e_0 Q_x^{\alpha\beta}-2(e_0-e_d)y^2[I_0(2x)-1],
     \label{eq:Simulation MDI_Q QE}
\end{align}
where
\begin{align}
    Q_C=&2(1-p_d)^2e^{-\mu^\prime/2}(1-(1-p_d)e^{-\eta\alpha/2})\\&\times(1-(1-p_d)e^{-\eta\beta/2}),\notag\\
    Q_E=&2p_d(1-p_d)^2e^{-\mu^\prime/2}[I_0(2x)-(1-p_d)e^{-\mu^\prime/2}],
\end{align}
\(\mu^\prime=\eta(\alpha+\beta)\), \(y=(1-p_d)e^{-\mu^\prime/4}\) and \(I_0(x)\) is the modified Bessel function of the first kind.} Using the above values as our observables, we can use the techniques introduced in \cref{Key rate analysis} and \cref{Analytical estimation} to bound the key rate. {Note that, in the numerical results presented, we account for the ten highest order terms in \cref{eq:Protocol Description_pseudo Poisson distribution} and \cref{eq:BB84 key rate anlaysis_fidelity between alpha1 and alpha2}, which can be shown to be of sufficiently high precision for the regimes of operation considered.}

\begin{figure}
    \centering
    \includegraphics[width=0.9\linewidth]{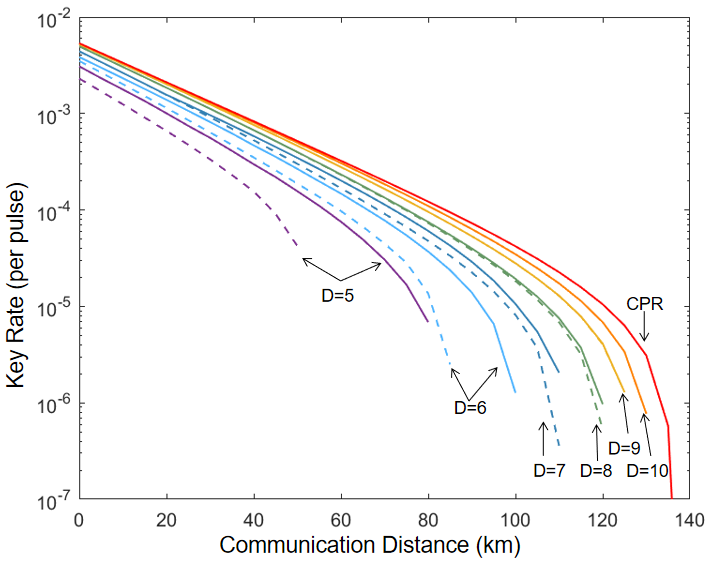}
    \caption{Secret key generation rates for the DPR BB84 protocol using numerical (solid lines) and analytical (dashed) methods. The mean number of photons for the signal and decoy states are optimized to get the maximum key rate. The best performing curve is based on the vacuum+weak decoy states and CPR. The remaining curves are generated when considering $D$ phase slices and vacuum+weak decoy states.}
    \label{fig:DPR BB84 using numerical and analytical methods}
\end{figure}
\begin{figure}
    \centering
    \includegraphics[width=0.9\linewidth]{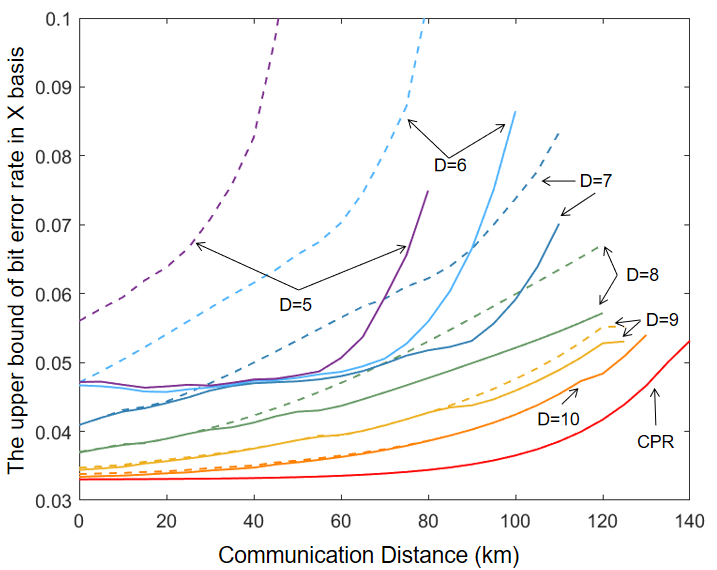}
    \caption{The upper bound \(e_{x,1}^{b,\mu U}\) on the bit error rate for the DPR BB84 protocol using numerical (solid) and analytical (dashed) methods. Results for the CPR case are shown as well. }
    \label{fig:bit error rate in DPR BB84}
\end{figure}
\begin{figure}
    \centering
    \includegraphics[width=0.9\linewidth]{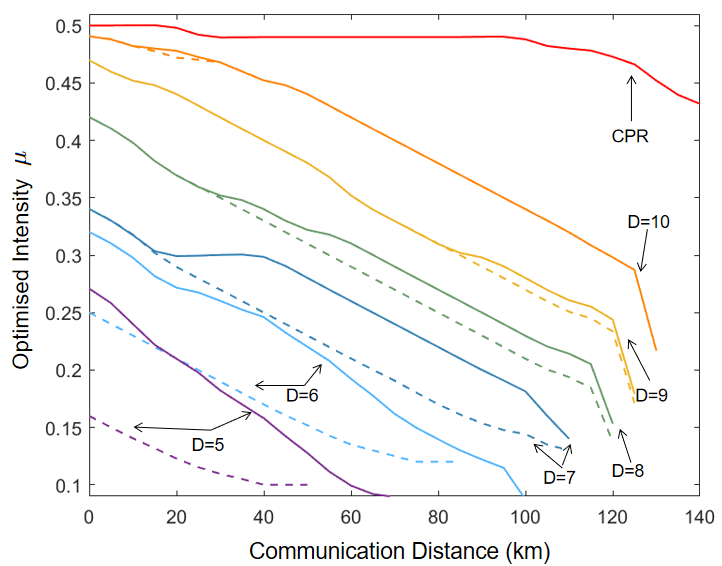}
    \caption{The optimized value of \(\mu\) for different number of phase slices ranging from \(D=5\) to \(D=10\), versus distance in the DPR BB84 protocol. A CPR related curve is provided for comparison. The numerical results are given in solid lines while analytical results are given in dashed lines.}
    \label{fig:optimized mu in DPR BB84}
\end{figure}

\begin{table}[!t]
\caption{Parameter Values Used in Simulations\label{table:system_parameters}}
\centering
\begin{tabular}{|c|c|}
\hline
Detection Efficiency & 0.045 \\
\hline
 Dark Count Rate Per Pulse, $p_d$& $1.7 \times 10^{-6}$ \\
\hline
Error Correction Inefficiency, $f$ & 1.16 \\
\hline
Phase Misalignment Error Rate, $e_d$  & 0.033 \\
\hline
Loss Factor & 0.20 dB/km\\
 \hline
\end{tabular}
\end{table}

\par \Cref{fig:DPR BB84 using numerical and analytical methods} presents the secret key rates for the vacuum+weak decoy BB84 protocol, using the numerical method (solid lines) described in \cref{BB84 QKD key rate analysis}, by solving the optimization problem in \cref{Box:NonlinearOptimizationBB84}, Approximate Formulation, and the analytical method (dashed lines) outlined in \cref{Analytical BB84}. The corresponding {upper bound on the $X$-basis bit error rate, \(e_{x,1}^{b,\mu U}\),} is shown in \cref{fig:bit error rate in DPR BB84}. 
The simulation parameters are shown in \cref{table:system_parameters}  \cite{gobby2004quantum,cao2015discrete}. During the optimization process for both numerical and analytical methods, the intensity \(\mu\) is varied within the range \([0, 0.5]\), and the decoy intensity \(\nu\) within \([0, 0.02]\), to optimize the key rate. The optimized values of \(\mu\) are displayed in \cref{fig:optimized mu in DPR BB84}. 
\par As shown in \cref{fig:DPR BB84 using numerical and analytical methods}, our analytical key rates exhibit increasing accuracy with larger values of the phase slice parameter \(D\) when compared with numerical results. When \(D > 7\), the analytical results become nearly indistinguishable from those obtained via numerical optimization, indicating that the analytical approximation is sufficiently tight in this regime. This is because when \(D\) increases, we get closer to the CPR case, where there is little difference between numerical and analytical results. Conversely, for smaller values of \(D\), the protocol remains functional but achieves secure key exchange only over shorter distances, highlighting the trade-off between implementation simplicity and performance, due to looser bounds on some parameters, such as \(e^{b,\mu}_{x,1}\) shown in \cref{fig:bit error rate in DPR BB84}.

The {upper bound} results on the $X$-basis bit error rate, presented in \cref{fig:bit error rate in DPR BB84}, can help explain the observed differences in key rate performance. The bit error rate \(e^{b,\mu}_{x,1}\) determines the bound on \(e^{p,\mu}_{z,1}\) via \cref{eq:BB84_bound between ep1 and eb1}. The bound on \(e^{b,\mu}_{x,1}\) is obtained via \cref{eq:BB84 Parameter_bound of y or eb with different intensities} and \cref{eq:BB84 key rate anlaysis_fidelity between alpha1 and alpha2}. From \cref{eq:BB84 key rate anlaysis_fidelity between alpha1 and alpha2}, we find that the fidelity $F_{\nu0}$ decreases as \(D\) decreases, leading to a looser bound in \cref{eq:BB84 Parameter_bound of y or eb with different intensities} for \(e^{b,\mu}_{x,1}\). Although both methods use the same constraints, numerical methods optimize over the entire range of variables while analytical methods directly pick the worst case for some individual variables, which leads to looser upper bounds in the analytical methods than in the numerical results. For sufficiently large values of  \(D\), the analytical bounds closely follow the numerical bounds across the entire communication distance. However, as the number of phase slices decreases, the error rates increase rapidly, and the gap between the analytical and numerical bounds becomes more pronounced. This behavior highlights the impact of phase slice resolution on the accuracy of the analytical estimation.

\par The performance of the DPR BB84 protocol is also closely tied to the choice of signal intensity \(\mu\), which directly influences both the secret key rate and the bit error rate. In \cref{fig:optimized mu in DPR BB84}, when the number of phase slices is \(D = 10\), the optimized intensity \(\mu\), obtained through numerical methods, starts at approximately 0.49. 
As the number of phase slices decreases or the communication distance increases, the optimized \(\mu\) correspondingly decreases in order to maintain a positive secret key rate.
This behavior can be explained as follows. When \(D\) is small, the phase-randomized states become more distinguishable, which compromises the fidelity between the prepared states. To ensure the fidelity in \cref{eq:BB84 Source_mixed states of xy fidelity} remains high, we must reduce \(\mu\); otherwise, the estimation of the phase error rate \(e^{p,\mu}_{z,1}\) in \cref{eq:BB84_bound between ep1 and eb1} becomes unreliable. Similarly, at longer communication distances, the channel loss increases significantly, which leads to greater information leakage or error correction cost, quantified by the term \(fQ_z^\mu h_2(E_z^\mu)\) in \cref{eq:BB84 Protocol_RDBB84}. To mitigate this, the system must again reduce \(\mu\), both to suppress multi-photon emissions and to preserve a positive key rate under adverse channel conditions.
\begin{figure}
    \centering
    \includegraphics[width=0.9\linewidth]{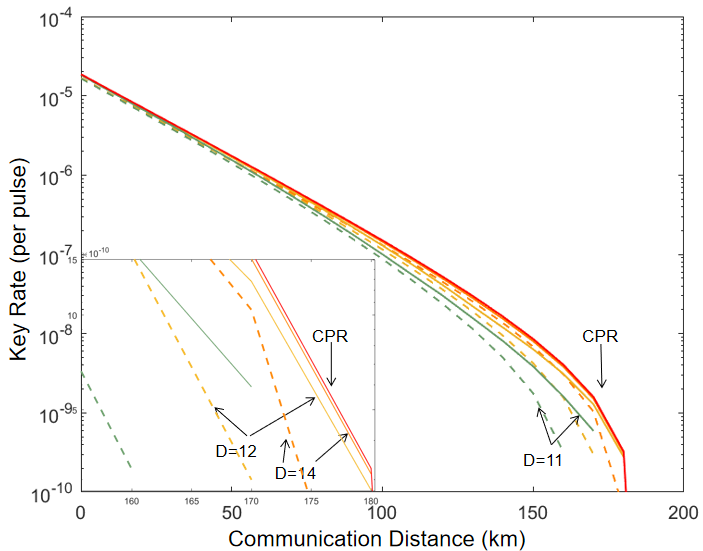}
    \caption{Secret key generation rates for the MDI-QKD protocol using numerical (solid lines) and analytical (dashed) methods. The mean number of photons for the signal and decoy states are optimized to get the maximum key rate. The best performing curve is based on the vacuum+weak decoy states and CPR. The remaining curves are generated when considering $D$ phase slices and vacuum+weak decoy states.}
    \label{fig:DPR MDI with Numerical and Analytical Results}
\end{figure}
\begin{figure}
    \centering
    \includegraphics[width=0.9\linewidth]{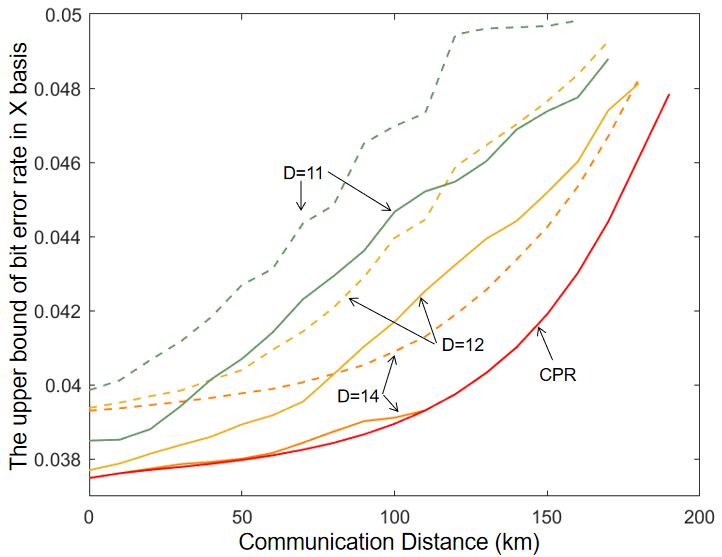}
    \caption{The upper bound on the bit error rate \(e_{x,11}^{b,\mu\mu}\), for the DPR MDI QKD protocol using numerical (solid) and analytical (dashed) methods. Results for the CPR case are shown as well.}
    \label{fig:bit error rate in DPR MDI}
\end{figure}
\begin{figure}
    \centering
    \includegraphics[width=0.9\linewidth]{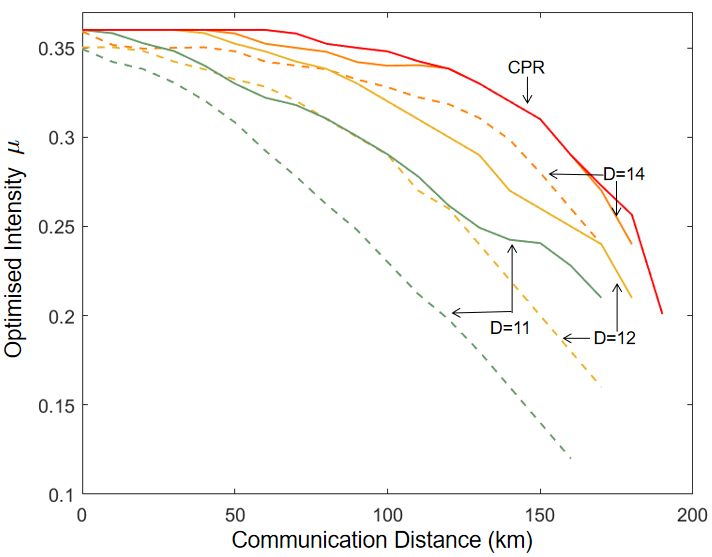}
    \caption{The optimum intensity \(\mu\) for different values of \(D\) versus distance in the DPR MDI QKD protocol.  The numerical results are shown in solid lines while analytical results are in dashed lines.}
    \label{fig:optimized mu in MDI}
\end{figure}

\par \Cref{fig:DPR MDI with Numerical and Analytical Results} shows the secret key generation rates for the DPR MDI QKD using the vacuum+weak decoy state technique. The corresponding {upper bound} on  \(e_{x,11}^{b,\mu\mu}\) is shown in \cref{fig:bit error rate in DPR MDI} and the optimized \(\mu\) is depicted in \cref{fig:optimized mu in MDI}. The key rates are computed using two approaches: the numerical optimization method presented in \cref{MDI QKD key rate analysis}, \cref{Box:NonlinearOptimizationMDI}, Approximate Formulation (solid lines), and our analytically derived solutions described in \cref{Analytical MDI} (dashed lines).
The simulation parameters are shown in \cref{table:system_parameters}. To determine the optimal key rate, we optimize the intensity \(\mu\) within the range \([0, 0.4]\) and the decoy intensity \(\nu\) within \([0, 0.02]\).

\par As shown in \cref{fig:DPR MDI with Numerical and Analytical Results}, once the number of discrete phases \(D\) reaches 14, the performance gap between the DPR protocol and the fully randomized case becomes negligible, which is consistent with the result in \cite{cao2020discrete}. Our analytical solutions exhibit strong agreement with the numerically optimized results for both the CPR and DPR cases with phase slice numbers \(D>10\).
{Note that, in \cref{fig:DPR MDI with Numerical and Analytical Results}, we need more phase slices for MDI QKD than for BB84 to approach the performance offered by CPR. This is due to two main factors. First, in MDI QKD, it is necessary to ensure that the fidelity of the states in \cref{eq:MDI Source_mixed states of xy fidelity} from both parties approaches unity, which demands more phase slices compared to the single-side fidelity in \cref{eq:BB84 Source_mixed states of xy fidelity}. Secondly, in \cref{eq:MDI Parameter_bound of y or eb with different intensities}, obtaining a tight bound requires additional phase slices because the expression involves two \(\epsilon\) terms, whereas \cref{eq:BB84 Parameter_bound of y or eb with different intensities} contains only one.}

\par \cref{fig:bit error rate in DPR MDI} shows that, when $D$ is sufficiently large, our analytical results for the {upper bounds on} the bit error rate closely match that of the numerical ones. The discrepancy observed at 0~km is attributed to the formulation in \cref{eq:MDI Parameter_bound of y or eb with different intensities}, where two \(\epsilon\) terms contribute to our analytical bound. Although both methods are subject to the same constraints, analytical bounds always select the worst-case value for each variable independently. In contrast, numerical bounds can become tight when the full set of variables is jointly considered. This discrepancy leads to an initial mismatch.

\par We also compare the optimized \(\mu\) with respect to different phase slices and communication distances in \cref{fig:optimized mu in MDI}. As shown in \cref{fig:optimized mu in MDI}, the maximum value of the optimized signal intensity \(\mu\) is approximately 0.37 in the CPR case. This value gradually decreases as the number of phase slices is reduced or as the communication distance increases, which is caused by the same reasons as we discussed in \cref{fig:optimized mu in DPR BB84}. 

Overall, in all results presented, we observe a speed up in processing time, when we use our analytical bounds. Although numerical optimization provides accurate performance estimates, it is computationally intensive and may hinder real-time parameter estimation. In contrast, our analytical bounds offer a fast and practical alternative for determining system parameters and the required level of privacy amplification.
\section{Summary \label{Summary}}
In this work, we presented an analytical framework for evaluating the performance of QKD protocols based on discrete-phase randomization (DPR). Existing DPR-based security proofs often rely on computationally intensive numerical optimization \cite{cao2015discrete,cao2020discrete}, which limits their practicality in real-time or resource-constrained environments. We derived analytical bounds for the secret key rates of BB84 and MDI QKD protocols under DPR. These bounds closely matched numerical results and significantly reduced processing time, particularly when the number of phase slices was sufficiently large. The proposed methods substantially reduce computational overhead and are applicable to other MDI-type protocols.


\par There are several directions to pursue future work. First, we can extend this framework to include finite-key analysis. Preliminary studies on DPR BB84 \cite{jin2024finite} and twin-field QKD \cite{lucamarini2018overcoming, wang2023finite} use semidefinite programming to obtain numerical bounds. Our analytical approach could make such analysis more tractable. Additionally, our techniques can be applied to the mode-pairing QKD protocol \cite{zeng2022mode}, which is another member of the MDI QKD family with improved scaling. {Another valuable direction for future work is extending our DPR analytical analysis to fully-passive QKD architectures \cite{wang2023fully,lu2023experimental}. Unlike DPR, where the phase is drawn from a known discrete set, fully-passive sources generate continuous, spontaneously-emitted phases whose probability distribution is not analytically characterized. This makes it difficult to construct the DPR decomposition and the associated closed-form bounds. Progress in this direction would require explicit models for the intrinsic phase distribution and for the conditional statistics with respect to post-selection. Establishing such models could enable computationally efficient analytical security bounds for fully-passive QKD systems.}

\section*{Acknowledgments}
\par The authors would like to thank M. Ghalaii, L. Wooltorton, G. Currás Lorenzo and X. Sixto for enlightening discussions. 

\appendices
\crefalias{section}{appendixsection}

\section{Analytical BB84 Bounds \label{Appendix: analytical BB84}}
\par In this appendix, we derive \cref{eq:Analytical BB84_Lower bound of Y1mu} and \cref{eq:Analytical BB84_Upper bound of W1mu} in the main text. We use \(\gamma\) to represent either \(X\) or \(Z\) basis.
\par First, we can rewrite \(Q_\gamma^\alpha\) and \(Q_\gamma^\alpha E_\gamma^\alpha\) as  \cite{ma2005practical}
\begin{equation}
    \begin{split}
        e^\alpha Q_\gamma^{\alpha}&=e^\alpha\sum_{k=0}^{D-1}p_k^\alpha Y_{\gamma,k}^{\alpha},\\
        e^{\alpha} Q_\gamma^\alpha E_\gamma^{\alpha}&=e^\alpha\sum_{k=0}^{D-1}p_k^{\alpha} W_{\gamma,k}^{\alpha},
    \end{split}
    \label{eq:Analytical BB84_ealbe Qalbe QEalbe}
\end{equation}
where \(e^\alpha = \exp(\alpha)\).  To bound \(Y_{\gamma,1}^\mu\), we use the above gain equations for two different intensity values, \(\alpha_1>\alpha_2\), to obtain
\begin{align}
         &e^{\alpha_1} Q_\gamma^{\alpha_1}-e^{\alpha_2} Q_\gamma^{\alpha_2}\notag\\
        =&e^{\alpha_1}p_0^{\alpha_1}Y_{\gamma,0}^{\alpha_1}-e^{\alpha_2}p_0^{\alpha_2}Y_{\gamma,0}^{\alpha_2}+e^{\alpha_1}p_1^{\alpha_1}Y_{\gamma,1}^{\alpha_1}-e^{\alpha_2}p_1^{\alpha_2}Y_{\gamma,1}^{\alpha_2}\notag\\&+\sum_{k=2}^{D-1}(e^{\alpha_1}p_k^{\alpha_1}Y_{\gamma,k}^{\alpha_1}-e^{\alpha_2}p_k^{\alpha_2}Y_{\gamma,k}^{\alpha_2})\notag\\
\leq&(e^{\alpha_1}p_0^{\alpha_1}-e^{\alpha_2}p_0^{\alpha_2})Y_{\gamma,0}^{0}+e^{\alpha_1}p_0^{\alpha_1}\epsilon_{\alpha_1 0}+e^{\alpha_2}p_0^{\alpha_2}\epsilon_{\alpha_2 0}\notag\\&+(e^{\alpha_1}p_1^{\alpha_1}-e^{\alpha_2}p_1^{\alpha_2})Y_{\gamma,1}^{\mu}+e^{\alpha_1}p_1^{\alpha_1}\epsilon_{\mu\alpha_1}+e^{\alpha_2}p_1^{\alpha_2}\epsilon_{\mu\alpha_2}\notag\\&+\sum_{k=2}^{D-1}[(e^{\alpha_1}p_k^{\alpha_1}-e^{\alpha_2}p_k^{\alpha_2})Y_{\gamma,k}^{\mu}+e^{\alpha_1}p_k^{\alpha_1}\epsilon_{\mu\alpha_1}+e^{\alpha_2}p_k^{\alpha_2}\epsilon_{\mu\alpha_2}]\notag\\
\leq&(e^{\alpha_1}p_0^{\alpha_1}-e^{\alpha_2}p_0^{\alpha_2})Q_\gamma^0+(e^{\alpha_1}p_1^{\alpha_1}-e^{\alpha_2}p_1^{\alpha_2})Y_{\gamma,1}^{\mu}\notag\\&+A_{\alpha_1\alpha_2}\sum_{k=2}^{D-1}[e^\mu p_k^{\mu}Y_{\gamma,k}^{\mu}+\frac{1}{A_{\alpha_1\alpha_2}}(e^{\alpha_1}p_k^{\alpha_1}\epsilon_{\mu\alpha_1}+e^{\alpha_2}p_k^{\alpha_2}\epsilon_{\mu\alpha_2})]\notag\\&+e^{\alpha_1}p_0^{\alpha_1}\epsilon_{\alpha_1 0}+e^{\alpha_2}p_0^{\alpha_2}\epsilon_{\alpha_2 0}+ e^{\alpha_1}p_1^{\alpha_1}\epsilon_{\mu\alpha_1}+e^{\alpha_2}p_1^{\alpha_2}\epsilon_{\mu\alpha_2}\notag\\
=&(e^{\alpha_1}p_0^{\alpha_1}-e^{\alpha_2}p_0^{\alpha_2})Q_\gamma^0+(e^{\alpha_1}p_1^{\alpha_1}-e^{\alpha_2}p_1^{\alpha_2})Y_{\gamma,1}^{\mu}\notag\\&+A_{\alpha_1\alpha_2}[e^{\mu}Q^\mu_\gamma-e^\mu p_0^\mu Y_{\gamma,0}^\mu-e^\mu p_1^\mu Y_{\gamma,1}^\mu]\notag\\&+e^{\alpha_1}p_0^{\alpha_1}\epsilon_{\alpha_1 0}+e^{\alpha_2}p_0^{\alpha_2}\epsilon_{\alpha_2 0}\notag\\&+(e^{\alpha_1}-e^{\alpha_1}p_0^{\alpha_1})\epsilon_{\mu\alpha_1}+(e^{\alpha_2}-e^{\alpha_2}p_0^{\alpha_2})\epsilon_{\mu\alpha_2}\notag\\
\leq&(e^{\alpha_1}p_0^{\alpha_1}-e^{\alpha_2}p_0^{\alpha_2}-A_{\alpha_1\alpha_2}e^\mu p_0^\mu)Q_\gamma^0\notag\\&+(e^{\alpha_1}p_1^{\alpha_1}-e^{\alpha_2}p_1^{\alpha_2}-A_{\alpha_1\alpha_2}e^\mu p_1^\mu)Y_{\gamma,1}^{\mu}\notag\\&+ A_{\alpha_1\alpha_2}e^{\mu}Q_\gamma^\mu+A_{\alpha_1\alpha_2}e^\mu p_0^\mu \epsilon_{\mu 0}+e^{\alpha_1}p_0^{\alpha_1}\epsilon_{\alpha_1 0}+e^{\alpha_2}p_0^{\alpha_2}\epsilon_{\alpha_2 0}\notag\\&+(e^{\alpha_1}-e^{\alpha_1}p_0^{\alpha_1})\epsilon_{\mu\alpha_1}+(e^{\alpha_2}-e^{\alpha_2}p_0^{\alpha_2})\epsilon_{\mu\alpha_2}
    \label{eq:Appendix BB84_Analytical bound calculation for Y1mu}
\end{align}
where {\(p_k^\alpha\) is the probability distribution function defined in \cref{eq:Protocol Description_pseudo Poisson distribution}, \(\epsilon\) terms are given by \cref{eq:BB84 Parameter_bound of y or eb with different intensities}, and \(A_{\alpha_1\alpha_2}\) is a constant defined in \cref{eq:Appendix_BB84_Parameter_A}.} In the first inequality, we use \cref{eq:BB84 Parameter_bound of y or eb with different intensities} repeatedly to write all yield terms either as a function of intensity \(\mu\) or \(0\). Note that the left hand side of this inequality is a function of observable parameters. In the next two inequalities, we have written the right hand side as a function of $Y_{\gamma,1}^\mu$, the parameter we want to bound, as well as other observables and constants. Consequently, the lower bound of  \(Y_{\gamma,1}^\mu\) can be obtained as shown in \cref{eq:Analytical BB84_Lower bound of Y1mu}.
\par For the parameters related to errors, the upper bounds are derived similarly. To derive \(W_{x,1}^{\mu U}\), we can use the following bound:{
\begin{align}
    Q_x^{\alpha_1}E_x^{\alpha_1}    &=p_0^{\alpha_1}W_{x,0}^{\alpha_1}+p_1^{\alpha_1}W_{x,1}^{\alpha_1}+\sum_{k=2}^{D-1}p_k^{\alpha_1}W_{x,k}^{\alpha_1}\notag\\
    &\geq p_0^{\alpha_1}(W_{x,0}^{0}-\epsilon_{\alpha_1 0})+p_1^{\alpha_1}(W_{x,1}^{\mu}-\epsilon_{\mu\alpha_1)})\notag\\    &=p_0^{\alpha_1}Q_{x,0}^0E_{x,0}^0+p_1^{\alpha_1}W_{x,1}^\mu-p_0^{\alpha_1}\epsilon_{\alpha_1 0}-p_1^{\alpha_1}\epsilon_{\mu\alpha_1}.
\end{align}
}Consequently, we can get an upper bound on \(W_{x,1}^\mu\) as given by \cref{eq:Analytical BB84_Upper bound of W1mu}. 

\section{Analytical MDI QKD Bounds \label{Appendix: analtycial MDI}}
\par In this appendix, we derive \cref{eq:Analytical MDI_Y11L} and \cref{eq:Analytical MDI bounds_W11U} in the main text. We set \(\alpha=\beta=\nu\) in analysis and use \(\gamma\) to represent either \(Z\) or \(X\) basis.
\par First, we can rewrite \(Q_\gamma^{\alpha\beta}\) and \(Q_\gamma^{\alpha\beta}E_\gamma^{\alpha\beta}\) as  \cite{ma2005practical}
\begin{align}
    e^{\alpha+\beta} Q_\gamma^{\alpha\beta} &= e^{\alpha+\beta} \sum_{k,l=0}^{D-1} p_k^\alpha p_l^\beta Y_{\gamma,kl}^{\alpha\beta}, \notag \\
    e^{\alpha+\beta} Q_\gamma^{\alpha\beta}E_\gamma^{\alpha\beta} &= e^{\alpha+\beta} \sum_{k,l=0}^{D-1} p_k^\alpha p_l^\beta e_{\gamma,kl}^{b,\alpha\beta} Y_{\gamma,kl}^{\alpha\beta}.
    \label{eq:Analytical MDI_Discrete ealbe Qalbe2}
\end{align}
\par In order to use other observables to bound \(Y_{\gamma,11}^{\mu\mu}\), \(Q_\gamma^{\alpha\beta}\) can be expanded as follows:
\begin{align}
        &e^{\alpha+\beta} Q_\gamma^{\alpha\beta}\notag\\=&\sum_{k,l=0}^{D-1}e^{\alpha+\beta}p_k^{\alpha}p_l^{\beta}Y_{\gamma,kl}^{\alpha\beta}\notag\\
        =&S_{1}+S_{2}+S_{3}+S(\alpha,\beta).
    \label{eq:Appendix MDI_Qalpbet 2}
\end{align}
where
\begin{align}
        S_{1}=&e^{\alpha+\beta}\sum_{l=0}^{D-1}p_0^{\alpha}p_l^{\beta}Y_{\gamma,0l}^{\alpha\beta},\notag\\
        S_{2}=&e^{\alpha+\beta}(p_1^{\alpha}p_0^{\beta}Y_{\gamma,10}^{\alpha\beta}+\sum_{k=2}^{D-1}p_k^{\alpha}p_0^{\beta}Y_{\gamma,k0}^{\alpha\beta}),\notag\\
        S_{3}=& e^{\alpha+\beta}p_1^{\alpha}p_1^{\beta}Y_{\gamma,11}^{\alpha\beta},\notag\\
        S(\alpha,\beta)=&e^{\alpha+\beta}(\sum_{k=2}^{D-1}p_k^{\alpha}p_1^{\beta}Y_{\gamma,k1}^{\alpha\beta}+\sum_{l=2}^{D-1}p_1^{\alpha}p_l^{\beta}Y_{\gamma,1l}^{\alpha\beta}\notag\\&+\sum_{k=2}^{D-1}\sum_{l=2}^{D-1}p_k^{\alpha}p_l^{\beta}Y_{\gamma,kl}^{\alpha\beta}).
    \end{align}
\(S_{1}, S_{2}\) and \(S_{3}\) can be bounded separately as follows:
\begin{align}
        S_{1} \leq&e^{\alpha+\beta}p_0^{\alpha}\sum_{l=0}^{D-1}p_0^{0}p_l^{\beta}(Y_{\gamma,0l}^{0\beta}+\epsilon_{\alpha 0})=e^{\alpha+\beta}p_0^{\alpha}(Q_\gamma^{0\beta}+\epsilon_{\alpha 0}),\notag\\
        S_{2}=&e^{\alpha+\beta}(\sum_{k=0}^{D-1}p_k^{\alpha}p_0^{\beta}Y_{\gamma,k0}^{\alpha\beta}-p_0^{\alpha}p_0^{\beta}Y_{\gamma,00}^{\alpha\beta})\notag\\
        \leq&e^{\alpha+\beta}(p_0^{\beta}\sum_{k=0}^{D-1}p_k^{\alpha}(Y_{\gamma,k0}^{\alpha0}+\epsilon_{\beta 0})-p_0^{\alpha}p_0^{\beta}(Y_{\gamma,00}^{00}-\epsilon_{\alpha 0}-\epsilon_{\beta0}))\notag\\
        =&e^{\alpha+\beta}(p_0^{\beta} Q_\gamma^{\alpha 0}-p_0^{\alpha}p_0^{\beta}Q_\gamma^{00}+(p_0^{\beta} +p_0^{\alpha}p_0^{\beta})\epsilon_{\beta 0}+p_0^{\alpha}p_0^{\beta}\epsilon_{\alpha 0}),\notag\\
        S_{3}\leq&e^{\alpha+\beta} p_1^{\alpha}p_1^{\beta}(Y_{\gamma,11}^{\mu\mu}+\epsilon_{\alpha\mu}+\epsilon_{\beta\mu}).
    \label{eq:Appendix MDI_Four parts of Qalpbet}
\end{align}
The three inequalities in \cref{eq:Appendix MDI_Four parts of Qalpbet} are due to \cref{eq:MDI Parameter_bound of y or eb with different intensities}. Using \cref{eq:Appendix MDI_Qalpbet 2} and \cref{eq:Appendix MDI_Four parts of Qalpbet}, an upper bound on \(Q_\gamma^{\alpha\beta}\) can be expressed as
\begin{align}
        &e^{\alpha+\beta} Q_\gamma^{\alpha\beta}\notag\\\leq&e^{\alpha+\beta}(p_0^{\alpha}Q_\gamma^{0\beta}+p_0^{\beta} Q_\gamma^{\alpha 0}-p_0^{\alpha}p_0^{\beta}Q_\gamma^{00}+p_1^{\alpha}p_1^{\beta}Y_{\gamma,11}^{\mu\mu}\notag\\&+(p_0^{\alpha}+p_0^{\alpha}p_0^{\beta})\epsilon_{\alpha 0}+(p_0^{\beta}+p_0^{\alpha}p_0^{\beta})\epsilon_{\beta 0}\notag\\&+p_1^{\alpha}p_1^{\beta}(\epsilon_{\alpha\mu}+\epsilon_{\beta\mu}))+S(\alpha,\beta)\notag\\
    =&T_1+ e^{\alpha+\beta}p_1^{\alpha}p_1^{\beta}Y_{\gamma,11}^{\mu\mu}+S(\alpha,\beta)+\epsilon_{\bar{s}},
    \label{eq:Appendix MDI_UUQ bound with S}
\end{align}
where $T_1$ and $ \epsilon_{\bar{s}}$ are defined in \cref{eq:Analytical MDI_Discrete ealbe Qalbe} and \cref{eq:MDI Analytical Bound_epsilons} separately.
\par Till now, \(Q_\gamma^{\alpha\beta}\) has been almost replaced by a sum of other observables and parameters that can be calculated directly, except \(Y_{\gamma,11}^{\mu\mu}\), which is the variable we need in the key rate formula, and \(S(\alpha,\beta)\). Therefore, the next task is to bound \(S(\alpha,\beta)\) with some observables to provide a bound for \(Y_{\gamma,11}^{\mu\mu}\). Based on \cref{eq:MDI Parameter_bound of y or eb with different intensities}, we get a bound for \(S(\alpha,\beta)\) as follows:
\begin{align}
    &S(\alpha,\beta)\notag
\\\leq&e^{\alpha+\beta}(\sum_{k=2}^{D-1}p_k^\alpha p_1^\beta (Y_{\gamma,k1}^{\mu\mu}+\epsilon_{\alpha\mu}+\epsilon_{\beta\mu})\notag
\\&+\sum_{l=2}^{D-1}p_1^\alpha p_l^\beta (Y_{\gamma,1l}^{\mu\mu}+\epsilon_{\alpha\mu}+\epsilon_{\beta\mu})\notag
\\&+\sum_{k=2}^{D-1}\sum_{l=2}^{D-1}p_k^\alpha p_l^\beta (Y_{\gamma,kl}^{\mu\mu}+\epsilon_{\alpha\mu}+\epsilon_{\beta\mu}))\notag
\\
    \leq&e^{2\mu}(A\sum_{k=2}^{D-1}p_k^\mu p_1^\mu Y_{\gamma,k1}^{\mu\mu}+B\sum_{l=2}^{D-1}p_1^\mu p_l^\mu Y_{\gamma,1l}^{\mu\mu}\notag
\\&+C\sum_{k=2}^{D-1}\sum_{l=2}^{D-1}p_k^\mu p_l^\mu Y_{\gamma,kl}^{\mu\mu})+e^{\alpha+\beta}(\sum_{k=2}^{D-1}p_k^\alpha p_1^\beta(\epsilon_{\alpha\mu}+\epsilon_{\beta\mu})\notag
\\&+\sum_{l=2}^{D-1}p_1^\alpha p_l^\beta (\epsilon_{\alpha\mu}+\epsilon_{\beta\mu})+\sum_{k=2}^{D-1}\sum_{l=2}^{D-1}p_k^\alpha p_l^\beta (\epsilon_{\alpha\mu}+\epsilon_{\beta\mu}))\notag
\\
    \leq&G\cdot S(\mu,\mu)+\epsilon_{s}\notag
\\
    \leq&Ge^{2\mu} (Q_\gamma^{\mu\mu}-p_0^{\mu}(Q_\gamma^{0\mu}+ Q_\gamma^{\mu 0})+p_0^{\mu}p_0^{\mu}Q_\gamma^{00}\notag
\\&-p_1^{\mu}p_1^{\mu}Y_{\gamma,11}^{\mu\mu}+2(p_0^{\mu} +p_0^{\mu}p_0^{\mu})\epsilon_{\mu 0})+\epsilon_{s}\notag
\\
    =&GT_s-Ge^{2\mu}p_1^{\mu}p_1^{\mu}Y_{\gamma,11}^{\mu\mu}+2Ge^{2\mu}(p_0^{\mu}+p_0^{\mu}p_0^{\mu})\epsilon_{\mu 0}+\epsilon_{s}\notag
\\
    =&GT_s-Ge^{2\mu}p_1^{\mu}p_1^{\mu}Y_{\gamma,11}^{\mu\mu}+2G\epsilon_{T_s}+\epsilon_{s},
\label{eq:Appendix MDI_S bounds}
\end{align}
where {\(\epsilon\) terms are given by \cref{eq:MDI Parameter_bound of y or eb with different intensities,eq:MDI Analytical Bound_epsilons}, and \(A,B,C\) are constants defined in \cref{eq:Analytical MDI_Discrete bounds A B C}.} 

Finally, using \cref{eq:Appendix MDI_UUQ bound with S} and \cref{eq:Appendix MDI_S bounds}, the lower bound of \(Y_{\gamma,11}^{\mu\mu}\) is obtained in \cref{eq:Analytical MDI_Y11L}.
\par In terms of \(e^{p,\mu\mu}_{z,11}\) in the key rate formula, we first bound \(e^{b,\mu\mu}_{x,11}\) based on \(W_{x,11}^{\mu\mu}\). Similar to what we did in \cref{eq:Appendix MDI_UUQ bound with S}, we replace all the \(Q\) terms with \(QE\) terms, change the less than or equal sign to the greater than or equal sign, and change the signs of \(\epsilon\) terms from plus signs to minus signs. Consequently, a lower bound on \(Q_x^{\alpha\beta}E_x^{\alpha\beta}\) is given by
\begin{align}     &e^{\alpha+\beta}Q_x^{\alpha\beta}E_x^{\alpha\beta}\notag
\\\geq&e^{\alpha+\beta}(p_0^{\alpha}Q_x^{0\beta}E_x^{0\beta}+p_0^{\beta} Q_x^{\alpha 0}E_x^{\alpha 0}-p_0^{\alpha}p_0^{\beta}Q_x^{00}E_x^{00}\notag\\&+p_1^{\alpha}p_1^{\beta}W_{x,11}^{\mu\mu}-(p_0^{\alpha}+p_0^{\alpha}p_0^{\beta})\epsilon_{\alpha 0}-(p_0^{\beta}+p_0^{\alpha}p_0^{\beta})\epsilon_{\beta 0}\notag
\\&-p_1^{\alpha}p_1^{\beta}(\epsilon_{\alpha\mu}+\epsilon_{\beta\mu}))+S(\alpha,\beta)\notag
\\
    \geq&T_2+ e^{\alpha+\beta}p_1^{\alpha}p_1^{\beta}W_{x,11}^{\mu\mu}-\epsilon_{\bar{s}},
    \label{eq:Appendix MDI_UUQE bound}
    \end{align}
{where \(T_2\) and \(\epsilon_{\bar{s}}\) are given in \cref{eq:MDI Analytical Bound_T2} and \cref{eq:MDI Analytical Bound_epsilons}, respectively}. The upper bound on \(W_{x,11}^{\mu\mu}\) can then be calculated as in \cref{eq:Analytical MDI bounds_W11U}.

\bibliographystyle{IEEEtran}
\bibliography{references}

\end{document}